\input{aipcheck}

\documentclass[
    ,final            
  ,  sort&compress               
  ]
  {aipproc}

\usepackage{amsmath}

\layoutstyle{6x9}

\begin{document}

\def\r{{\mathbf r}}
\def\freq{\omega_0}

\title{Radiation of a circulating quark in strongly coupled ${\cal N}=4$ super Yang-Mills theory\footnote{Talk presented by D. Nickel at QCD@Work, June 20-23rd, 2010, Martina Franca, Italy.}}


\author{Christiana Athanasiou}{
  address={Center for Theoretical Physics, MIT, Cambridge, MA 02139, USA}
}

\author{Paul M. Chesler}{
  address={Center for Theoretical Physics, MIT, Cambridge, MA 02139, USA}
}

\author{Hong Liu}{
  address={Center for Theoretical Physics, MIT, Cambridge, MA 02139, USA}
}

\author{Dominik Nickel}{
  address={Institute for Nuclear Theory, University of Washington, Seattle, WA 98195, USA}
}

\author{Krishna Rajagopal}{
  address={Center for Theoretical Physics, MIT, Cambridge, MA 02139, USA}
}

\begin{abstract}
The energy density and angular distribution of power radiated by a quark undergoing circular motion in strongly coupled ${\cal N}=4$ supersymmetric Yang-Mills (SYM) theory is computed using  gauge/gravity duality.
The results are qualitatively similar to that of synchrotron radiation produced by an electron in circular motion in classical electrodynamics: At large velocities the quark emits radiation in a narrow beam along its velocity vector with a characteristic opening angle $\alpha \sim 1/\gamma$ and radial thickness scaling like $\sim 1/\gamma^3$.  
\end{abstract}

\maketitle


\section{Introduction and Motivation}

\noindent

\vspace{-320pt}
\hspace{11.5cm}
MIT-CTP 4177
\vspace{307pt}

\vspace{-307pt}
\hspace{11.18cm}
INT-PUB-10-042 
\vspace{294pt}

\noindent
Through the discovery of gauge/gravity duality~\cite{Maldacena:1997re}, the dynamics of quantum fields in certain strongly coupled gauge theories can be addressed by the dynamics of classical fields living in a higher dimensional spacetime. In particular  strongly coupled real-time phenomena in ${\cal N}=4$ super Yang-Mills (SYM) theory have become accessible to reliable calculation for the limit of large colors $N_{\rm c}$ and large but finite t'Hooft coupling $\lambda$. Much of the analysis in this direction has so far been focused on finite temperature systems, whose strongly coupled plasmas are expected to behave similarly to the strongly coupled quark-gluon plasma of QCD being produced and studied in heavy ion collision experiments at the Relativistic Heavy Ion Collider~\cite{Adcox:2004mh,Back:2004je,Arsene:2004fa,Adams:2005dq}. For reasons that will be outlined below, we will however consider the specific case of a test quark in circular motion in the vacuum of SYM theory, i.e. synchrotron radiation. For this case we are able to find an analytic result that can be confronted with expectations from the weakly coupled regime, electrodynamics and a parton picture.

Because the vacuum of SYM theory is not similar to that of our world, studying the radiation of a test quark in circular motion in this theory does not have direct phenomenological motivation. However, we shall find that the angular distribution of the radiation is very similar to that of classic synchrotron radiation: when the quark is moving along its circle with an increasingly  relativistic velocity, the radiation is produced in an increasingly tightly collimated beam --- a beam which, if Fourier analyzed, contains radiation at increasingly short wavelengths and high frequencies.

First, this finding is interesting in its own terms. From the point of view of the nonabelian gauge theory it is not obvious why the angular distribution of the radiation at strong coupling is similar to what is known for the weakly coupled regime.
Indeed a study related to zero temperature radiation analyzed decays of off-shell bosons in strongly coupled holographic conformal field theories~\cite{Hofman:2008ar} and it was shown that there is no correlation between the boson's  spin and the angular distribution of power radiated through a sphere at infinity. In other words, if one prepares a state containing an off-shell boson with definite spin, in the rest frame of the boson the event averaged angular distribution of power at infinity will always be isotropic.  
Similar behavior regarding the isotropization of radiation at strong  coupling was also reported in Refs.~\cite{Hatta:2008tx,Chesler:2008wd}.
A reasonable question is whether the isotropization is a characteristic of the particular initial states studied in Ref.~\cite{Hofman:2008ar} or whether it ia a natural process which happens during the propagation of strongly coupled radiation.
It has been suggested  in Ref.~\cite{Hatta:2008tx} that the mechanism responsible for the isotropization is that 
parton branching is not suppressed at strong coupling and, correspondingly, successive branchings can scramble any initially preferred direction in the radiation as it propagates out to spatial infinity.
If the intuition about the emission and propagation of radiation derived from these calculations applies to radiation from an accelerated test charge, our result should show isotropization of the energy density when propagating towards infinity.
However, this is not found.

Second, our results in this formal setting nevertheless open the way to new means of modeling jet quenching in heavy ion collisions. If we were to add a nonzero temperature to our calculation, we could watch the tightly collimated beam of synchrotron radiation interact with the strongly coupled plasma that would then be present.  The beam of radiation should be slowed down from the speed of light to the speed of sound and should ultimately thermalize, and it would be possible to study how the length- and time-scales for these processes depend on the wavelength and frequency of the beam.  
The authors of Ref.~\cite{Fadafan:2008bq} have shown that the power that it takes to move the quark in a circle is given by the vacuum result~\cite{Mikhailov:2003er} even for a test quark moving through the strongly coupled plasma present at finite temperature $T$ as long as $\omega_0^2 \gamma^3 \gg \pi^2 T^2$. Here $\omega_0$ is the angular frequency, $v$ the quark's speed and $\gamma=1/\sqrt{1-v^2}$.
In the opposite regime, where $\omega_0^2 \gamma^3 \ll \pi^2 T^2$, the power it takes to move the quark in a circle through the plasma is the same as it would be to move the quark in a straight line.  This power  has been computed using gauge/gravity duality in Refs.~\cite{Herzog:2006gh,Gubser:2006bz}, and corresponds to pushing the quark against a drag force.  The analogue of our calculation --- namely following where the dissipated or ``radiated'' power goes --- was done, again using gauge/gravity duality, in
Refs.~\cite{Gubser:2007xz,Chesler:2007an, Gubser:2007ga,Chesler:2007sv}. It was found that the radiation pattern eventually takes the form of hydrodynamic excitations at distances large compared to $1/T$.
For an extension of the present calculation to nonzero temperature, we therefore expect synchrotron radiation for $\omega_0^2 \gamma^3 \gg \pi^2 T^2$ at length scales that are small compared to $1/T$, which should eventually thermalize locally, likely converting into an outgoing hydrodynamic wave moving at the speed of sound, before ultimately dissipating and thermalizing completely.
Watching these processes occur may give insight into jet quenching in the strongly coupled quark-gluon plasma of QCD.

\section{Gravitational Setup}

\begin{figure}[t]
\includegraphics[scale=0.45]{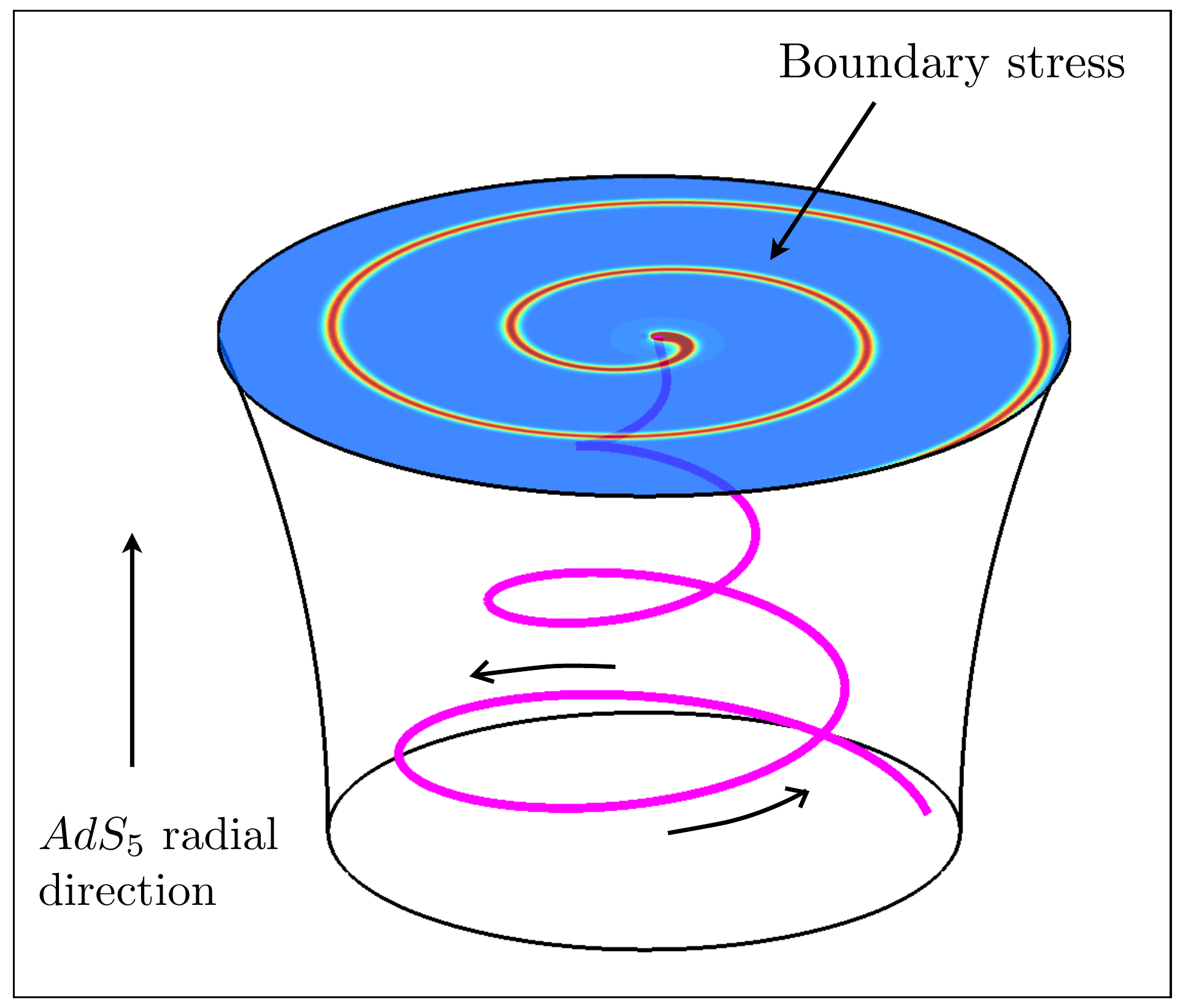}
\caption  {\label{bulk2boundary}
Cartoon of the gravitational description of synchrotron radiation at strong coupling: The $5d$ geometry of AdS$_5$ is shown as a foliation with leaves isomorphic to Minkowski space and understood as slices in the figure. The boundary, as an example, is mapped out with color-coding indicating the induced energy density at a given time.
The string (purple line) is attached to the boundary corresponding to the location of the quark in the dual theory and hangs down into the bulk.
}
\end{figure}

The determination of the energy radiated by a moving test quark (specifically, an infinitely massive spin $1/2$-particle of a ${\cal N}=2$ hypermultiplet) coupled to SYM theory in the large $N_{\rm c}$ limit for large but finite $\lambda$ can be translated into a classical gravitational description in an AdS$_5$ background using the gauge/gravity dictionary \cite{Maldacena:1997re, Witten:1998qj}.
A cartoon of the the dual realization is shown in Fig.~\ref{bulk2boundary}:
The four dimensional boundary of AdS$_5$ has the geometry of ordinary Minkowski space. Ending at the boundary is a classical string whose endpoint follows a trajectory that corresponds to the trajectory of the quark in the dual field theory. The specific condition in our setup that the endpoint rotates about a specific (Minkowski space) axis results in the entire string rotating about the same axis with the string coiling around on itself over and over.
Since we consider a single test quark in the large $N_{\rm c}$ limit, we can neglect the backreaction of the string profile onto the geometry. However, the rotating motion of the string creates gravitational radiation which propagates up to the boundary of the geometry. Just like electromagnetic fields induce surface currents on conductors, the gravitational disturbance near the boundary induces a $4d$ stress tensor on the boundary \cite{Brown:1992br, deHaro:2000xn}, which corresponds to  the expectation value of the stress tensor in the dual quantum field theory. 
So, by doing a classical gravitational calculation in five dimensions (which happens to be somewhat analogous to a classical electromagnetic calculation) we can compute the pattern of radiation in the boundary quantum field theory, including all quantum effects and working in a strong coupling regime in which quantum effects can be expected to dominate.
As indicated in Fig.~\ref{bulk2boundary}, the induced $4d$ stress tensor inherits the coiled structure of the rotating string.

In the following we will use the convention
\begin{equation}
\label{metric}
ds^2 = \frac{L^2}{u^2}\left [-dt^2 + d \r^2 + du^2 \right ]\,,
\end{equation}
for the metric of the AdS$_5$ spacetime,  where $L$ is the $AdS_5$ curvature radius, $u$ is the (inverse) $AdS_5$ radial coordinate with the boundary located at $u = 0$.
We first outline the determination of the string profile, which then serves as an input for the calculation of the induced energy pattern. For details we refer to Ref.~\cite{Athanasiou:2010pv}.

\section{The rotating string}
\label{RotatingStringSection}

The dynamics of classical strings are governed by the Nambu-Goto action
\begin{equation}
S_{\rm NG} = - \frac{\sqrt{\lambda}}{{2 \pi L^2} }\int d \tau \,d \sigma \sqrt{-g}\,,
\label{SNG}
\end{equation}
where $\sigma$ and $\tau$ are the coordinates on the worldsheet of the string, and $g= {\rm det}\, g_{ab}$ where $g_{ab}$ is the induced worldsheet metric. The string profile is given by an embedding $X^M(\tau,\sigma)$ of the worldsheet into AdS$_5$. 
As we are interested in quarks which rotate at constant frequency $\freq$ about the $\hat z$ axis, it is convenient to choose $\tau = t$ and $\sigma = u$. Furthermore the possible set of solutions can be constraint to the parameterization
\begin{equation}
X^{M}(t,u) = (t,\r_s(t,u),u)\,,
\label{Xstring}
\end{equation}
where in spherical coordinates $\{r,\theta,\varphi\}$ the three-vector
$\r_s$ is given by
\begin{equation}
\r_s(t,u) \equiv \left (R(u), {\textstyle \frac{\pi}{2}},\phi(u) +\freq t \right ).
\label{StringWorldsheetParametrization}
\end{equation}
The Nambu-Goto action then simplifies to
\begin{equation}
\label{NB2}
S_{\rm NG} = - \frac{\sqrt{\lambda}}{{2 \pi}} \int dt \,du \, \mathcal
L\,,
\end{equation}
where
\begin{equation}
\mathcal L = \frac{\sqrt{(1-\freq^2 R^2 )(1 + R'^2) + R^2 \phi'^2}}
{u^2}\,.
\label{NGLagrangian}
\end{equation}
To determine the shape of the string in Fig.~\ref{bulk2boundary}, we
must extremize Eq.~(\ref{NB2}) in $R(u)$ and $\phi(u)$.

For real-valued profile parameterizations $R(u)$ and $\phi(u)$ , it turns out that the solutions for the corresponding Euler-Lagrange equations are uniquely given by~\cite{Athanasiou:2010pv}
\begin{equation}
R(u) = \sqrt{v^2 \gamma^2 u^2 + R_0^2}\,,
\quad
\phi(u) = - u \gamma \freq + \arctan\left(u \gamma \freq\right)
\label{RSolution}
\end{equation}
where $v = R_0 \freq$ is the velocity of the quark
and $\gamma = 1/\sqrt{1-v^2}$. 

As a by-product the shape of the rotating string also determines the total power radiated, as we now explain.
The string has an energy density $\pi_0^0$ and energy flux $\pi_0^1$, where
\begin{equation}
\pi_M^a \equiv  \frac{\delta S_{\rm NG}}
{\delta (\partial_a X^M)}
\end{equation}
with $S_{\rm NG}$ given in (\ref{SNG}) and where energy conservation on the string worldsheet requires $\partial_a \pi^a_M=0$.
The energy flux down the rotating string then turns out to be
\begin{equation}
\label{stringflux}
P=-\pi^1_0 = \frac{\sqrt{\lambda}}{2 \pi} a^2\, ,
\end{equation}
where  $a = v\omega_0 \gamma^2$ is the proper acceleration of the test quark.
The energy flux down the string is the energy that must be supplied by the external agent that moves the test quark in a circle; by energy conservation, therefore, it is the same as the total power radiated by the quark and agrees with the findings in Refs.~\cite{Fadafan:2008bq,Mikhailov:2003er}.

\section{Gravitational perturbation and induced energy density}

The starting point for the determination of the induced stress-energy tensor are the linearized Einstein equations
\begin{align}
    &- D^2 \, h_{MN} +2 D^{P} D_{(M} h_{N) P}
      -D_{M}D_{N} \, h +\frac{8}{L^2} \, h_{MN}
\nonumber
\\ \label{lin1} &{}
    + \left (D^2 h -D^{P} D^{Q} \, h_{P Q} -\frac{4 }{L^2} \, h \right)
G_{MN}^{(0)}  =  2 \kappa_5^2 \; t_{MN} \, ,
\end{align}
where the full metric $G_{MN}= G^{(0)}_{MN} +h_{MN}$ is linearized around the unperturbed metric $G^{(0)}_{MN}$ defined in Eq.~(\ref{metric}), $h \equiv h^{M}_{\ M}$, $D_M$ is the covariant derivative with respect to $G^{(0)}_{MN}$, $\kappa_5^2$ is the $5d$ gravitational constant and $t_{MN}$ is the $5d$ stress
tensor of the string.

According to the gauge/gravity dictionary, the on-shell gravitational action
\begin{equation}
S_G = \frac{1}{2\kappa_5^2}\int d^5 x \sqrt{-G}\left({\cal R}+\frac{12}{L^2}\right) + S_{GH}
\end{equation}
is the generating functional for the boundary stress
tensor~\cite{Witten:1998qj, deHaro:2000xn}.
Here, $G$ is the determinant of $G_{MN}$, ${\cal R}$ is its Ricci scalar, and $S_{GH}$ is the Gibbons-Hawking boundary
term~\cite{Gibbons:1976ue}, discussed and evaluated in the present context in Ref.~\cite{Chesler:2007sv}.
The source of the dual stress-energy tensor is identified with
\begin{equation}
g_{\mu \nu}(x) \equiv \lim_{u \to 0} \frac{u^2}{L^2} G_{\mu \nu}(x,u)\,,
\label{BoundaryMetric}
\end{equation}
such that the boundary stress tensor is then given by \cite{deHaro:2000xn}
\begin{equation}
\label{stress}
T^{\mu \nu}(x) = \frac{2}{\sqrt{-g}} \frac{\delta S_{\rm G}}{\delta g_{\mu \nu}(x)}\,,
\end{equation}
with $g$ denoting the determinant of $g_{\mu \nu}$.  The gravitational perturbations are sourced by the $5d$ string stress
tensor, for which in general
\begin{equation}
t^{MN}
=
-\frac{T_0}{\sqrt{-G}}\sqrt{-g}g^{ab}
\partial_a X^M \partial_b X^N
\delta^{3}(\r-\r_s)\,.
\end{equation}

Focusing on the energy density $\mathcal{E}(t,\r)\equiv T^{00}(t,\r)$, it is then possible -- though non-trivial -- to derive~\cite{Athanasiou:2010pv}
\begin{eqnarray}
\mathcal{E}(t,\r) &=&
\frac{L^3}{\pi}
\int d^4r'
\int_0^\infty \!\! du \,\,
\theta(t-t')
\Big[
\left(
4u t_{00}
-
t_{M5}\nabla'_M \mathcal{W}
\right)
\frac{\delta''(\mathcal{W})}{u^2}
\nonumber\\
&&
+\vert \r-\r' \vert^2
\left(
4t_{00}-4t_{55}+2t_{ii}
\right)
\frac{\delta'''(\mathcal{W})}{3u}
-
\left(
t_{ij}\nabla'_i \mathcal{W} \nabla'_j \mathcal{W}
\right)
\frac{\delta'''(\mathcal{W})}{2u}
\Big]
\,,
\end{eqnarray}
where
\begin{eqnarray}
\mathcal W  &\equiv&  -(t-t')^2 + u^2 +|\r-\r'|^2
\end{eqnarray}
and $\nabla'_M$ is the partial derivative with respect to $X'^M=\{t',\r',u\}$.
Because we have analytic expressions for the profile of the rotating string, it turns out that we can furthermore evaluate ${\cal E}$ explicitly, albeit in a tedious calculation for which we again refer to Ref.~\cite{Athanasiou:2010pv}. We then arrive at the central result of this work:
\begin{eqnarray}
{\mathcal E}(t,\r)
&=& \frac{\sqrt{\lambda}}{24 \pi^2 \gamma^4 r^6 \Xi^6}
\Big[
4 r \gamma^2 \Xi  (t_{\rm ret}{-}t) 
+
7 r \gamma^2\freq^2\Xi (t_{\rm ret}{-}t)^3 
+
4 \gamma^2\freq^2 (t_{\rm ret}{-}t)^4
\\ \nonumber
&&
\phantom{\frac{\sqrt{\lambda}}{24 \pi^2 \gamma^4 r^6 \Xi^6}}
+
(2 \gamma^2 {-} 4 r^2 v^2 \gamma^2 \freq^2 \sin^2\theta{+}3 r^2 \gamma^4\freq^2\Xi^2)(t_{\rm ret}{-}t)^2
-
2 r^2 \Xi ^2
\\ \nonumber
&&
\phantom{\frac{\sqrt{\lambda}}{24 \pi^2 \gamma^4 r^6 \Xi^6}}
+
8v\gamma^2\freq r   (t_{\rm ret}{-}t)   (t_{\rm ret}{-}t { + }r \, \Xi )\sin \theta   \cos (\varphi {-} \freq t_{\rm ret}  )
\Big]\,,\label{eq:Ex}
\end{eqnarray}
where
\begin{equation}
\label{xieq}
\Xi(t,\r) \equiv  \frac{| \r {-} \r_s(t,0) |    -    \r \cdot {\dot {\r}}_s(t,0)}{\vert\r\vert}\,,
\end{equation}
and
\begin{equation}
\label{tret0}
t - t_{\rm ret} - | \r - \r_s(t_{\rm ret},0)| = 0\,.
\end{equation}
We have been able to obtain an analytic expression for our result in terms of the retarded time $t_{\rm ret}$.  For the case of a charge in circular motion, $t_{\rm ret}$ must be evaluated numerically, similar to the corresponding case in classical electrodynamics~\cite{Jackson}.

Our results simplify in the far zone $\vert\r\vert\to\infty$. With $\xi=1-v\sin\theta\sin(\varphi-\omega_0t_{\rm ret})$, here we find
\begin{equation}
\label{farzoneeps}
\vert\r\vert^2{\mathcal E}(t,\r) =  
\frac{\freq^2 \sqrt{\lambda }}{24 \pi ^2 }
\frac{4-4v^2 \sin^2 \theta - 7 \xi + 3 \xi^2 \gamma^2}{\xi^6 \gamma^2}
\end{equation}
and from the continuity equation $\hat{\r}\cdot{\bf S}(t,\r)={\mathcal E}(t,\r)$ for the asymptotic energy flux.
This allows us to obtain  the time-averaged angular distribution of power
\begin{align}
\label{dpdOmega}
\frac{dP}{d \Omega}
&=
\frac{v^2 \gamma^3 \freq^2 \sqrt{\lambda}}{32 \pi^2}
\frac{ 
	5 \gamma^2-1- v^2 \gamma^2 \cos (2 \theta )
}{
  \left(\gamma ^2 \cos ^2 \theta +\sin ^2\theta \right)^{5/2}
}\,.
\end{align}
We observe that radiated power is focused about $\theta = \pi/2$ with a characteristic width $\sim 1/ \gamma$ and 
-- upon integrating over all solid angles -- we verify the total power radiated agrees with Eq.~(\ref{stringflux}).

\section{Illustration}
\label{discussion}

\begin{figure*}[t]
\includegraphics[scale=.315]{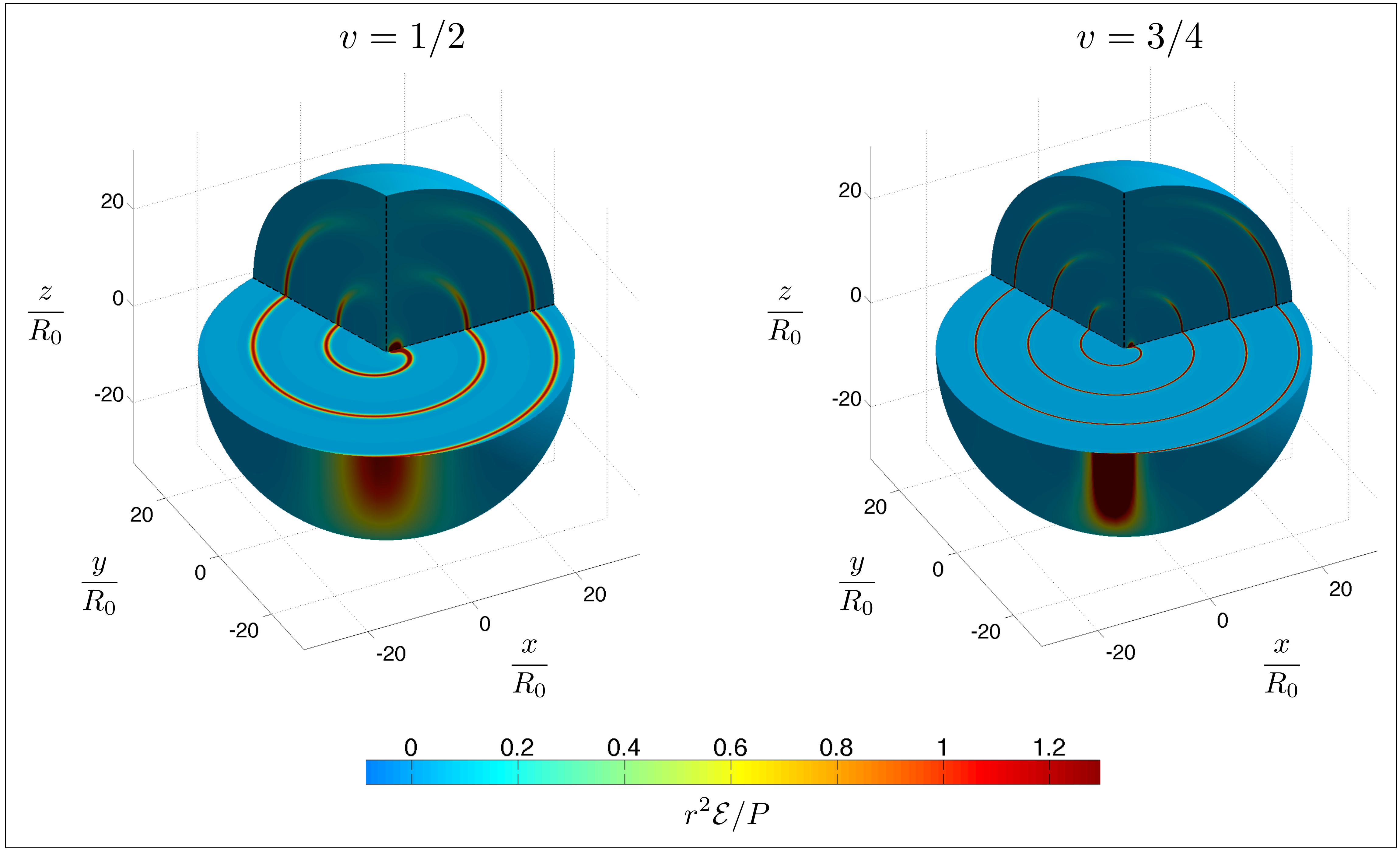}
\caption
    {\label{blingbling}
   Left: A cutaway plot of $r^2 \mathcal E/P$ for $v = 1/2$.
   Right: A cutaway plot of $r^2 \mathcal E/P$ for $v = 3/4$.
   In both plots the quark is at $x = R_0$, $y=0$ at the time shown and its 
   trajectory lies in the plane $z = 0$.  The cutaways coincide with the planes 
   $z = 0$, $\varphi = 0$ and $\varphi =  7 \pi / 5$.  At both velocities
   the energy radiated by the quark is concentrated along a spiral structure
   which propagates radially outwards at the speed of light.  The spiral is 
   localized about $\theta = \pi/2$ with a characteristic width $\delta \theta \sim 1/\gamma$.
   As $v \to 1$ the radial thickness $\Delta$ 
   of the spirals rapidly decreases like $\Delta \sim 1/\gamma^3$.  
   }
\end{figure*}

Fig.~\ref{blingbling} shows two cutaway plots of the energy density $\mathcal E$
at strong coupling (multiplied by $r^2/P$ where 
$P$ is the total power radiated) produced by a quark moving 
on a circle of radius $R_0$ at velocities $v = 1/2$ 
and $v = 3/4$.  The figure is obtained by evaluating (\ref{eq:Ex}) numerically.
The motion of the quark is confined to the plane $z = 0$ and 
at the time shown the quark is at $x = R_0$, $y = 0$, and is rotating 
counter clockwise. The cutaways in the plots show the energy density 
on the planes $z = 0$, $\varphi = 0$ and $\varphi = 7 \pi / 5$ where $\varphi$ is the azimuthal angle.   
As is evident from the figure, energy is radiated outwards in a spiral 
pattern as the quark accelerates along its trajectory.  This radiation falls off like $1/r^2$ and hence has a constant
amplitude in the figure and propagates radially outwards at the speed of light.   
The figure shows the location of the energy density at one time; as a function of time, the entire pattern of energy density rotates with constant angular frequency $\omega_0= v/R_0$. This rotation of the pattern is equivalent to propagation of the radiation outwards at the speed of light.
As seen by an observer far away from the quark, the radiation appears as a short pulse 
just like a rotating lighthouse beam does to a ship at sea.

From Fig.~\ref{blingbling} we also see that the outward going pulse of radiation that does not broaden as it propagates. This can be confirmed analytically.
As time progresses, the pulses move outward at the speed of light, and an observer at large $\vert\r\vert$ sees repeating flashes of radiation.
The figure underlines that even though the gauge theory is nonabelian and strongly coupled, the narrow pulses of radiation propagate outward without any hint of broadening or isotropization.   We find the same result at much larger values of $\gamma$, where the pulses become even narrower -- their widths are proportional to $1/\gamma^3$, as for classical electrodynamics~\cite{Jackson} and also as for weak coupling in SYM theory~\cite{Athanasiou:2010pv}.

Not visible in Fig.~\ref{blingbling}  but certainly worth mentioning is the fact that the energy density becomes negative before and after the pulse for sufficiently large velocities. We attribute this to quantum effects, showing that the classical gravitational setup is indeed dual to a strongly coupled quantum field theory.


\begin{theacknowledgments}
This research was supported in part by the DOE Offices of Nuclear and High Energy Physics under grants \#DE-FG02-94ER40818,  \#DE-FG02-05ER41360 and \#DE-FG02-00ER41132.
The work of HL is supported in part by a U.~S.~Department of Energy (DOE) Outstanding Junior Investigator award.
\end{theacknowledgments}



\bibliographystyle{aipproc}   
\bibliography{refs}

\end{document}